# Meteor Shower Detection with Density-Based Clustering

Glenn Sugar[1]*, Althea Moorhead[2], Peter Brown[3], and William Cooke[2]

[1]Department of Aeronautics and Astronautics, Stanford University, Stanford, CA 94305
[2]NASA Meteoroid Environment Office, Marshall Space Flight Center, Huntsville, AL, 35812
[3]Department of Physics and Astronomy, The University of Western Ontario, London N6A3K7, Canada
*Corresponding author, E-mail: gsugar@stanford.edu

**Abstract**

We present a new method to detect meteor showers using the Density-Based Spatial Clustering of Applications with Noise algorithm (DBSCAN; Ester et al. 1996). DBSCAN is a modern cluster detection algorithm that is well suited to the problem of extracting meteor showers from all-sky camera data because of its ability to efficiently extract clusters of different shapes and sizes from large datasets. We apply this shower detection algorithm on a dataset that contains 25,885 meteor trajectories and orbits obtained from the NASA All-Sky Fireball Network and the Southern Ontario Meteor Network (SOMN). Using a distance metric based on solar longitude, geocentric velocity, and Sun-centered ecliptic radiant, we find 25 strong cluster detections and 6 weak detections in the data, all of which are good matches to known showers. We include measurement errors in our analysis to quantify the reliability of cluster occurrence and the probability that each meteor belongs to a given cluster. We validate our method through false positive/negative analysis and with a comparison to an established shower detection algorithm.

## 1. Introduction

A meteor shower and its stream is implicitly defined to be a group of meteoroids moving in similar orbits sharing a common parentage. However, as meteoroids ejected from the same common parent necessarily have orbits which change with respect to one another over time due to differential planetary perturbations, the threshold at which a stream stops becoming a single defined entity is uncertain. As a result, it is unclear exactly how many meteor showers exist. There are frequent additions to the list of known showers as well as revisions to their characteristics. Multiple new showers were discovered in 2014 alone; for instance, Jenniskens (2014) reported 105 potential new meteor showers with an additional 23 new components to established showers; Segon et al. (2014) described a "set of 24 new showers"; and SonotaCo et al. (2014) reported the discovery of the yet another meteor shower, the April alpha Capricornids. The Meteor Data Center[1] (MDC) provides evidence of this changing catalog of meteor showers. As of August 27, 2015, there are 135 "pro tempore" showers from journal papers in submission. Some showers have multiple entries with different values for important parameters such as mean solar longitude and radiant location. These variations are sometimes due to real changes in meteor streams, but can also arise from differing biases between surveys and shower detection methods. It is therefore useful to have multiple methods for shower detection so that shower existence and properties can be independently confirmed.

---

[1] http://www.astro.amu.edu.pl/~jopek/MDC2007/Etc/streamworkingdata.csv

There are three main quantitative approaches used to extract showers from a set of meteoroid orbits or meteor radiants. The first involves comparing meteors to a parameter grid. Showers are detected as grid cells or points near which the density of meteor radiants is high. An example of this approach can be found in Younger et al. (2009), who detects 37 distinct showers by finding active regions in a gridded radiant space. Another example is seen in Brown et al. (2010b), who uses a 3D wavelet transform to detect 109 meteor showers in a dataset of over 3 million radar meteors. Molau (2007) uses another variant of this approach, searching for radiant density local maxima in a discrete probability space to detect showers.

The second approach compares meteors to known showers via a dissimilarity parameter (often called a *D*-parameter or *D*-criterion). Some examples of this method include Veres and Toth (2010) and Wu and Williams (1992), the latter who used it for the analysis of the Quadrantids. Galligan (2001) and Valsecchi et al. (1999) give an overview of various *D*-parameters used for shower detection. Meteors that have a *D*-parameter less than a predetermined critical value when compared to a shower will be associated with that shower. A major drawback for this method is that it prevents the discovery of new showers and relies on the assumption that the parameters of known showers are accurate. However, this method is computationally efficient, $O(n)$, since it does not require meteor-meteor comparisons.

Finally, the third approach compares meteors to meteors. Many studies have used this approach, including Jopek (1995), Jopek et al (2010), Rudawska et al. (2015), and Lindblad (1971). All of these used a version of this method that will associate two meteors with each other if their *D* value is less than a critical *D* value. A cluster is defined as a group where each member is associated with at least one of the other members, and showers are identified as clusters that meet certain requirements such as a minimum number of members. This serial association method is preferred for shower surveys over one that uses known shower mean orbits because it allows for the detection of new showers and does not rely on previous experiments that could have different observational biases and errors. The main drawback of this approach is the need to establish a false-positive association rate, a topic not normally addressed in contemporary meteor shower detection studies.

A single linkage meteor-meteor association method is a special case of the more general DBSCAN algorithm (Ester et al. 1996). Rather than always associating two points if they are within a specified distance of each other, DBSCAN allows the user to specify how many points must be within a specified distance of each other for an association to be made. This enforces a minimum cluster size. DBSCAN is also computationally fast, with an average runtime complexity of $O(n \log n)$. The scikit-learn Python package contains a DBSCAN function that is able to compute Euclidean norms very efficiently (Pedregosa et al. 2011). Therefore, there is an incentive to develop a *D*-parameter based on the Euclidean norm of a vector, which we describe in Section 2.1. Because of its ability to extract clusters of different shapes and sizes, the ability to set a minimum cluster size, and the lack of an a priori shower knowledge requirement, DBSCAN proved to be a useful algorithm for detecting meteor showers. Also, unlike other cluster detection algorithms such as k-means or mean-shift, DBSCAN does not require all points to be assigned to a cluster. This is a necessary attribute since the data will contain a significant amount of sporadic meteors that do not belong to any showers/clusters.

In this paper, we apply DBSCAN to data obtained from the NASA's All-Sky Fireball Network and the Southern Ontario Meteor Network. These camera networks provide nighttime observation of meteors over much of North America, detecting over 25,000 individual from January 2006 through August 2015 (Cooke and Moser 2012; Weryk et al. 2008). Our observations are complete to magnitude -4, but include meteors as faint as magnitude 0. The system records basic meteor characteristics and automatically calculates trajectories and orbital parameters with the ASGARD software (Brown et al. 2010a). This data set is particularly amenable to shower searches as the fraction of all meteors belonging to showers is believed to be larger in the centimeter-size range detected by these networks (Hughes, 1976).

## 2. Methods

This section describes our shower detection method used on the NASA and SOMN data set. We have applied some basic quality filters to our data set. First, we require that $Q_*$, the camera-meteor-camera angle from the ensemble of all detected cameras for a particular event, be greater than 15°. We also require that $v_g \leq 75$ km/s and that the associated uncertainty $\delta v_g \leq (0.1 \cdot v_g + 1)$ km/s, where $v_g$ is the meteor's geocentric velocity. These filters exclude the most poorly measured meteors from our dataset.

A meteor must be detected by multiple cameras within a network for it to be entered in a data set. However, even though the NASA and SOMN networks have overlapping fields of view, their camera stations do not communicate with each other. As a result, some meteors have duplicate entries in the combined NASA/SOMN data set. We must therefore remove these duplicated when merging the two databases. When two meteors appear in both networks and have less than 2 seconds difference in their detection time and less than 5° difference in their starting latitude and longitude, we consider them to be duplicates, and remove the data point with fewer detection stations. The resulting filtered dataset contains 25,885 meteors.

### 2.1 DBSCAN Vector

We use solar longitude ($\lambda_\odot$), geocentric ecliptic latitude of the radiant ($\beta_g$), Sun-centered ecliptic longitude of the radiant ($\lambda_g - \lambda_\odot$), and geocentric velocity ($v_g$) to generate a six dimensional vector (Equation 1). Note that $\lambda_g$ is the geocentric ecliptic longitude of the radiant. This six-dimensional vector automatically handles longitude wrapping; i.e., $\lambda_\odot = 1°$ is only 2° from $\lambda_\odot = 359°$ rather than 358°. We normalize the six parameters so that the influence of each variable in cluster determination is comparable.

$$\begin{bmatrix} \cos(\lambda_\odot) \\ \sin(\lambda_\odot) \\ \sin(\lambda_g - \lambda_\odot)\cos(\beta_g) \\ \cos(\lambda_g - \lambda_\odot)\cos(\beta_g) \\ \sin(\beta_g) \\ v_g \left(\frac{km}{s}\right) \Big/ 72\left(\frac{km}{s}\right) \end{bmatrix} \quad (1)$$

The first two elements give the meteor position because the meteoroid intersects the Earth's orbit at $\lambda_\odot$. The next three elements specify the unit vector opposite the direction of the meteor's velocity. The last element is the normalized geocentric velocity magnitude. The Euclidean norm of the difference between two meteor vectors is similar in concept to the Valsecchi et al. (1999) *D*-parameter. However, the norm has the benefit of fast computation time because of the built-in functionality of the scikit-learn Python package. All vector elements range from [-1, 1] except for the element dependent on velocity, which has a range of [0, 1]. Because the velocity measurement has the largest error, we allow the velocity term to carry less weight.

## 2.2 DBSCAN Parameters

DBSCAN classifies each meteor as a core, boundary, or noise point (Ester et al. 1996). A core point has a minimum number of points (including itself), *N*, within a specified hypersphere of radius ϵ centered on the core point. Boundary points are within ϵ of a core point, but there are fewer than *N* points in their ϵ-neighborhood. Finally, noise points are neither core nor boundary points. For our purposes, we define both core points and boundary points are cluster members and noise points are sporadic meteors. While unlikely, it is possible for a boundary point to be a member of multiple clusters. In this rare case, the point is randomly assigned to one of the potential clusters. This does not pose a major issue for our analysis because we run one thousand iterations of DBSCAN using the measurement errors reported by ASGARD, so a meteor that is on the border of two clusters will most likely be assigned to both.

DBSCAN requires only two user-defined parameters, ϵ and *N*, to sort a data set into clusters. We set *N*=5 in order to avoid reporting showers that consist of 4 or fewer members. There are generally two ways to calculate an appropriate ϵ. One can make a sorted nearest neighbor distance plot similar to Fig. 1 and either [1.] pick a point of high curvature or [2.] select the value corresponding to the expected percentage of clustered meteors. We opted to use the second method; through trial and error, we found that 23% of the observed meteors are associated with showers for the combined NASA and SOMN dataset. If the percentage is set too high, distinct clusters will merge together. If the percentage is too low, some known showers will fail to be detected and showers will appear to be less active. In the limit of setting this percentage to 100%, DBSCAN will place all meteors into a single cluster. If the percentage is set to 0%, all meteors would be classified as sporadic and none would belong to a cluster. It is reasonable to expect this percentage to be to be constant because neither the sporadic background nor total shower activity have large year-to-year variability. Therefore, the percentage of shower meteors does not need to be recalculated every time new data is added to the data set as long as the equipment used to gather the data remains the same.

The quantity ϵ is calculated from the percentage of expected shower meteors as follows. The distance between the fourth nearest neighbor is calculated for each meteor vector using the Euclidean norm. These distances are then sorted by magnitude, and ϵ is chosen to be the 23$^{rd}$ percentile distance, 0.373 using our six-dimensional parameter space (Fig. 1). Note that ϵ is near a point of high curvature, which is an indication of a good ϵ value. We use the fourth nearest neighbor distance because we require a minimum of 5 members for shower identification. Therefore, the distance to the fourth-nearest neighbor must be less than ϵ for a meteor to be core point of a shower.

## 2.3 Measurement Errors

In order to take measurement errors into account, we clone each meteor 1000 times, assuming Gaussian uncertainties in ecliptic radiant and velocity. These uncertainties are calculated by the ASGARD software, which assumes an uncertainty in speed equal to the standard deviation of all the individual station speeds about the average and a radiant error determined by the non-linear fitting procedure for each trajectory following Borovicka et al. (1995). We assume zero error in $\lambda_\odot$ because we measure the time of occurrence to much higher precision and accuracy than radiant and velocity. Cloning the meteors produces 1000 datasets, each of which we pass through DBSCAN, resulting in 1000 sets of clusters. We generate a single master list by merging clusters that share at least 50% of one cluster's members. We also throw out any clusters that appear in less than 100 of the 1000 runs in order to remove spurious clusters composed of poorly measured meteors. We can calculate the probability that a meteor is part of a cluster by dividing the number of times that meteor was identified as a member of a cluster by the total number of times the cluster was detected. However, as discussed in Section 4.2, it is very likely that a meteor is a shower member even if it is only classified as one in a small percentage of the 1000 iterations.

### 2.4 Shower Identification

We compare our master cluster list with the MDC's working list of showers, again using Equation 1 to quantify similarity. We also calculate a crude measure of confidence in the shower identification as follows. We filter the MDC list of 974 shower entries by removing the 29 showers that do not have values for Solar longitude, right ascension, declination, or geocentric velocity. For any shower with multiple entries, we use the mean of the reported values, resulting in a list of 629 unique showers. Then, for each cluster in our data set, we find the distance to the nearest neighboring MDC shower in the same six dimensional space we use for DBSCAN. We observe that a histogram of the negative log of this distance is similar to a logistic distribution (Fig. 2). Using the least-squares method, we determine the best-fit parameters of the logistic distribution. Our confidence in a match is the CDF value of the fitted logistic distribution at the -log distance between the cluster and its closest MDC match (column 2 of Table 1).

### 3. Results

Using the DBSCAN method, we find 23 clusters in the combined NASA and SOMN dataset that were detected in at least 50% of the cloned datasets (strong detections), and an additional 8 clusters detected between 10% and 50% of the datasets (weak detections). The distinction between strong and weak detections is defined by our later analysis of simulated sporadic meteor datasets described in Section 4.1, which shows that the sporadic meteor population could generate clusters in less than 50% of the cloned datasets. While it is surprising that clusters detected as infrequently as 12% of the iterations have a close match to a real shower, numerous runs of our analysis detected 31 clusters that match the same MDC showers in over 10% of the 1000 iterations. The actual datasets that are passed to DBSCAN change slightly between every analysis because they are generated randomly based on the observation errors of the NASA and SOMN dataset. While the total number of detected clusters often changes after every analysis, the same 31 clusters are extracted once a 10% existence threshold is applied to clusters. The name, shower match confidence, number of times the cluster was detected in the 1000 cloned data sets, number of meteor members, mean solar longitude, mean radiant location, and mean velocity of the 31 clusters are reported in Table 1. Note that Table 1 is divided into strong and weak detections, and standard deviations appear in parentheses.

## 3.1 Kappa Cygnids

All clusters in Table 1 have a confidence match > 0.98 except for one cluster matched to the kappa Cygnids with a confidence value of 0.9181 (KCG2). DBSCAN yields another cluster that is matched to the KCGs, but with 0.9974 confidence (KCG1). The KCGs are an interesting shower with seven distinct entries in the MDC, all with different shower parameters (Cook 1973; Jopek et al. 2003; Molau et al. 2002; SonotaCo 2009). Koseki (2014) also reports sub-streams in the vicinity of the KCGs. The wide spread in reported KCG parameters and the possibility of KCG sub-streams could explain the comparatively low matching confidence of 0.9181 as well as the duplicate matches. Also, the low confidence cluster is a very weak detection, being detected in less than 15% of the total DBSCAN iterations. Fig. 3 shows the radiant location and solar longitude for all meteors identified in the two KCG clusters, as well as the location of five KCG showers reported in the MDC. If the values for the two KCG sub-streams reported in Molau et al. (2012) are averaged together and compared to the KCG cluster with a low confidence match, the confidence value increases from 0.9181 to 0.9994. This suggests that the dataset is too small to resolve the individual KCG sub-streams detected by Molau et al. (2012), and our analysis is merging these sub-streams into a single cluster.

## 3.2 Taurids

The Northern Taurids (NTAs) are another interesting shower detected by our algorithm. Inspection of Fig. 4 shows two distinct sub-clusters within the set of meteors identified as NTAs, and it is clear that DBSCAN merged the Northern and Southern Taurid (STA) showers into a single cluster. Like the KCG shower, there are multiple entries for the STAs in the MDC, with right ascension values ranging from 30.9° to 50.1°, and declination values ranging from 8.1° to 13.4° (Porubcan and Kornos 2002; Jopek et al. 2003; Brown et al. 2008; SonotaCo 2009; Brown et al. 2010b). In our survey, meteors near the SonotaCo (2009) STA radiant were merged with the cluster identified as the Northern Taurids (NTA; see Fig. 4). This demonstrates that one of DBSCAN's strengths, the ability to extract clusters of various shapes, could actually be undesirable depending on the user's needs. Because the NTA and STA are close together in our six-dimensional DBSCAN space, there will be some meteors that fall within the small gap due to measurement errors. These poorly measured meteors will cause a link between the NTA and STA showers, causing them to merge into a single cluster. However, NTA and STA meteors do have similar orbits, velocities, and radiants, as well as a common origin; the only real difference between the North and South branches is whether the collision with the Earth occurs at the ascending or descending node of the stream orbit. Therefore, DBSCAN correctly identifies the NTAs and STAs as part of the same complex. This region of parameter space contains another shower called the Southern October delta Arietids (SOA), and sometimes meteors in this region are classified as SOAs rather than STAs (Sekanina 1976).

## 4. Validation

The agreement between our detected clusters and known showers in the MDC is one indication of our algorithm's validity. We also performed three additional validation tests. The first consists of running the algorithm on a set of simulated sporadic meteors, and the second involves a qualitative comparison of the rate of sporadic meteors when showers are present and absent. The final test is a comparison between the detected showers from our new DBSCAN method and an established method that utilizes a 3D wavelet transform to detect showers (Brown et al. 2010b).

## 4.1 Simulated Data

With the goal of estimating the number of false positive clusters produced by the sporadic background, we tested our method on a set of simulated sporadic meteors generated from radiant and velocity distributions in Campbell-Brown (2008) and Solar longitude distributions in Campbell-Brown and Jones (2006). The simulated radiants were sampled from Gaussian distributions with mean and standard deviation found in Table 1 of Campbell-Brown (2008); velocities were sampled from the geocentric velocity distributions given in Figures 11, 13, and 15 of Campbell-Brown (2008); and Solar longitudes were sampled from the averaged 2002-2005 data in Figures 11, 13, and 15 of Campbell-Brown and Jones (2006).

As described in Section 2.2, we found that assuming sporadic meteors comprised 77% of the NASA and SOMN dataset gave reasonable results for the DBSCAN analysis. Therefore, we simulated 19,931 meteors (77% the number of meteors in the NASA and SOMN dataset) belonging to the north apex, antihelion, and north toroidal sporadic sources; the south toroidal and helion sources are not visible to our optical meteor network. The south apex source is partially visible, but has reduced strength compared to the north apex source, and therefore it was omitted. The choice to model the false positive rate for clustering using three sporadic sources, rather than three plus a partial source, is a conservative one.

Our simulated data belong to the antihelion, north apex, and north toroidal sources, which comprise 42.6%, 30.9%, and 26.5% of the total number of radiants, respectively. These fractions match the relative raw observed strength of these sources reported in Table 1 of Campbell-Brown (2008). The Sun-centered ecliptic radiants for our simulated data also match the parameters reported in Figure 8 of Campbell-Brown (2008). We assigned errors to individual simulated meteors by sampling fitted probability density distributions (PDFs) to scaled errors in the NASA and SOMN data set. Fig. 5 shows the three different PDFs used for assigning errors to the simulated meteors: the logarithm of the error in geocentric latitude of the radiant ($\delta_\beta$), the logarithm of the geocentric longitude of the radiant error ($\delta_\lambda$) divided by $\delta_\beta$ multiplied by the cosine of $\beta_g$, and the geocentric velocity error scaled by $0.1 v_g + 1$. While the logarithms in the first two distributions were taken for plotting purposes, the scaling factors and cosine term are used for more appropriate error assignments. For example, if two meteors had the same uncertainty in the radiant location, but one was found with $\beta_g \approx 90°$ and the other with $\beta_g \approx 0°$, the $\beta_g \approx 90°$ would have a much larger $\delta_\lambda$.

We applied our DBSCAN analysis to this simulated dataset using the same ϵ and *N* for DBSCAN as was used for the NASA and SOMN data (0.0373 and 5 respectively). We generated 1000 clones for each simulated meteor assuming Gaussian errors and ran the 1000 sets through DBSCAN. Three clusters were detected in over 10% of the iterations (Table 2), but no clusters were detected in over 33%. It is surprising that one of the clusters has a large confidence match (0.9639) with the April chi Librids. This suggests that it is possible for sporadic meteors to form dense enough regions in our parameter space for DBSCAN to detect a cluster that is not associated with a real shower despite having a large confidence match value. However, these false showers will not be detected in the majority of the cloned datasets. Therefore, we establish a threshold of 50% above which we can be confident in a shower detection. We refer to shower detections above this threshold as "strong" detections. "Weak" detections refer to clusters that appear in 10-50 % of the cloned data sets. Because these weak detections may be false positives, they will require additional confirmation.

### 4.2 Sporadic Rates

We assess the false positive and negative rates of meteors being classified as shower members by inspecting the number of meteors observed as a function of solar longitude (Fig. 6 and Fig. 7). If our method generates many false positives, we would expect a drop in sporadic activity during times of high shower activity due to sporadic meteors being falsely identified as shower meteors. If there were many false negatives, we would expect a rise in the sporadic observations during times of high shower activity. Therefore, by comparing the number of shower and sporadic meteors over time, we can assess how robust a method is to false positives and negatives. Because we produce 1000 cloned data sets, it is possible for a single meteor to be classified as sporadic in one set and a shower member in another. Therefore, we can either perform a binary assignment or a weighted assignment. For example, if a meteor were a member of a cluster in 400 out of the 1000 cloned data sets, a binary assignment would classify the meteor as a 100% shower member, while the weighted assignment would classify it as 40% shower and 60% to the sporadic. Fig. 6 and Fig.7 show the activity of the shower and sporadic population for binary and weighted assignments respectively. Notice in Fig. 7 during periods of high shower activity, such as $\lambda_\odot \approx 139°$, there are peaks in the sporadic distribution that correspond to peaks in the shower distribution. However, using binary assignment in Fig. 6, the result shows evidence of many fewer negatives for shower association. Therefore, it is likely that a meteor is a member of a shower even if it is only identified as a cluster member in a small percentage of the total 1000 cloned data sets.

### 4.3 Wavelet Method Comparison

The final validation test is a comparison between our results and the 3D wavelet method described in Brown et al. (2010b). We applied the wavelet algorithm to our all-sky dataset, initially with fairly permissive wavelet peak thresholds. These permissive settings resulted in the identifications of 191 wavelet peaks, far more than detected via DBSCAN. However, 30 wavelet peaks had a wavelet coefficient of 20 or greater, and most of these resembled DBSCAN-detected clusters. Most DBSCAN clusters have a close analog within this list of strong wavelet peaks; however, there are a few noteworthy cases where DBSCAN detects showers that the wavelet method does not and vice versa.

DBSCAN produces four weakly detected clusters with no nearby wavelet showers above the wavelet coefficient threshold: the December kappa Draconids (DKD), upsilon Andromedids (UAN), c Andromedids (CAN), and the Aurigids (AUR). These are all relatively low activity showers that appear in less than 50% of all cloned data sets and contain 31 or less meteors, so it understandable that the wavelet peaks fell below the threshold. The DKD detection is supported by an independent method using various *D*-parameters in Moorhead (2016). Close inspection of Fig. 8 shows two wavelet peaks below the threshold close to the DKDs, suggesting that the wavelet method split the DKDs into two sub-streams with wavelet peaks below the threshold. The UANs, CANs, and AURs were not detected above the wavelet coefficient threshold of 20 nor could they be confirmed using the method of Moorhead (2016), suggesting they might be false detections by DBSCAN. However, there are good agreements between the clusters and the MDC matches (0.9998, 0.9978, and 0.9894) for the AURs, CANs, and UANs respectively. Also, there are wavelet peaks below our wavelet coefficient threshold near all three of these showers. These conflicting points support the "weak" detection classification for these showers.

The region around RA = 45°, Dec = 15° contains seven distinct wavelet showers, but only two DBSCAN showers (NTA and SOA). However, as demonstrated in Section 3.2 and shown in Fig. 4, there are numerous sub-streams in the Taurid/SOA complex. While DBSCAN will merge most these sub-streams together, the wavelet method will extract individual peaks and identify each sub-stream as a single shower. A similar effect is seen around the KCG/GDR complex. The wavelet method detects four distinct peaks, while DBSCAN reports only 3 clusters. More examples of DBSCAN merging distinct wavelet peaks occur around the PERs, JPEs, DLMs, CAPs, and HYDs.

Another difference between the two results is the wavelet detection of three small showers at high declinations (RA = 117.6°, Dec = 76.9°; RA = 175.0°, Dec = 78.2°; and RA = 358.3°, Dec = 75.4°). These showers have relatively low wavelet coefficients: 36.2, 33.6, and 20.1 respectively. Out of these three showers, only one has a close match to an MDC shower; the wavelet peak corresponding to RA = 175.0°, Dec = 78.2° is matched to the October Camelopardalids (OCTs). However, the match is relatively weak with a confidence of 0.9481 (all of the DBSCAN matched confidence values are greater than 0.98 except for one cluster matched to the KCGs with 0.9181 confidence). The simulated sporadic data set showed that it is possible for a false detection to have a confidence match value as large as 0.9639, so it is possible that this is a false detection by the wavelet method. Additionally, the OCTs are a relatively weak shower; SonataCo (2009) found that the OCTs are 0.28% as active as the PERs. Assuming a similar ratio for our dataset, we would expect to see approximately 12 OCT meteors. Only the CANs had fewer than 12 meteors in the DBSCAN results, so the OCTs may be slightly below the sensitivity limit of DBSCAN for the dataset. Because of the weak pair matching to the MDC showers and the low activity of the OCTs, it not surprising that DBSCAN would miss this shower.

The wavelet peak around RA = 358.3°, Dec = 75.4° is associated with the 2014 May Camelopardalid outburst, which is not included in the MDC but was both predicted and detected (Ye and Wiegert 2014, Brown 2014, and Campbell-Brown et al. 2016). This is therefore an example of a false negative by our DBSCAN-based algorithm. However, the third high declination wavelet peak around RA = 117.6°, Dec = 76.9° is not associated with any known shower. This is likely a false positive from the wavelet algorithm rather than a newly discovered meteor shower for two reasons. Unlike the 2014 May Camelopardalids, meteors near this wavelet peak span many years and many days within each year, making it visible across a wide range of longitudes. Second, our data set has a relatively small sample size of compared to those used in previous studies. Examples of meteor surveys conducted on data gathered from cameras similar to the ones used in the NASA and SOMN networks include SonataCo (2009), which contains 39,208 meteor orbits, and Rudawska et al. (2015), which uses the EDMOND 5.0 database containing 144,749 orbits. It is therefore unlikely that a shower with regular annual activity would be missed by these networks, and so we conclude that it is likely not a real detection.

## 5. Conclusion

We have demonstrated that the DBSCAN algorithm can be used to effectively and efficiently detect meteor showers within a set of meteor observations. We have extended the DBSCAN cluster detection algorithm to incorporate measurement errors by producing cloned datasets based on assuming Gaussian measurement errors. We also provide a confidence parameter for how well a cluster matches to any known shower. No a priori knowledge of showers is used in cluster detection; however, all of our detected clusters had close matches to showers in the MDC database. We estimate that no false clusters are detected above a cluster existence threshold of 50%, as evidenced by the lack of consistent cluster detection in the DBSCAN iterations of simulated sporadic sources as well as the strong matches to MDC showers in the DBSCAN analysis of the NASA and SOMN data. There is evidence for valid cluster detection for clusters found in less than 50% of the cloned data sets, but those clusters must be corroborated to ensure they are not false positives.  As shown by our simulated sporadic dataset, it is possible for a small minority of the iterations to have false positive clusters formed by sporadic meteors. Also, this method allows for a good characterization of shower activity by classifying a meteor as a shower member if it has been assigned to a shower in at least 1 of the 1000 cloned data sets.  Fig. 6 and Fig. 7 show that there are either few false positives and negatives for classifying a meteor as a shower member, or the number of false positives and negatives roughly cancel each other out.

One surprising result is that nearly all clusters detected in less than 50% of DBSCAN iterations are associated with real meteor showers reported in the MDC.  One might expect that if a cluster is not detected in more than half of the iterations, it is not a reliable detection.  However, multiple DBSCAN runs of 1000 dataset realizations each indicate that our method reliably detects eight showers between 10-50% of the total DBSCAN iterations, each of which have good matches to showers in the MDC catalog.  Thus, by taking measurement errors into account by generating multiple cloned datasets, it may be possible to detect more meteor showers than through just one iteration of the shower detection algorithm.

Our shower detection method found 25 strong and 6 weak cluster detections in the combined NASA and SOMN dataset.  While this is much less than the 629 unique showers with orbital parameters found in the MDC database, there are many reasons for not detecting a shower within our dataset.  Showers can have a steep population index and are not detectable in the all-sky regime.  Poor weather during active showers will cause missed meteor observations.  Showers that are active during the day can be detected by radars, but will fail to be detected by the NASA and SOMN networks.  Showers with low declination radiants may not be observable from the Northern Hemisphere and therefore will not be detected in our dataset.  Also, showers can be very weak and require many more orbits for detection.

We present the results of a DBSCAN-based shower identification algorithm applied to data from obtained from optical all-sky meteor cameras, but this method can be easily applied to data from other instruments such as radar or narrow-field cameras.  However, an appropriate prediction of the percentage of shower meteors must be used for each dataset.  For example, a radar meteor network sensitive to millimeter-sized meteoroids will detect proportionally more sporadic meteors.  Thus, the predicted percentage of shower meteors should be lower than the 23% used for our all-sky data.

## 6. Acknowledgements

This work was supported in part by NASA Cooperative Agreement NNX15AC94A and the NASA Marshall Space Flight Center's internship program.

**Figures**

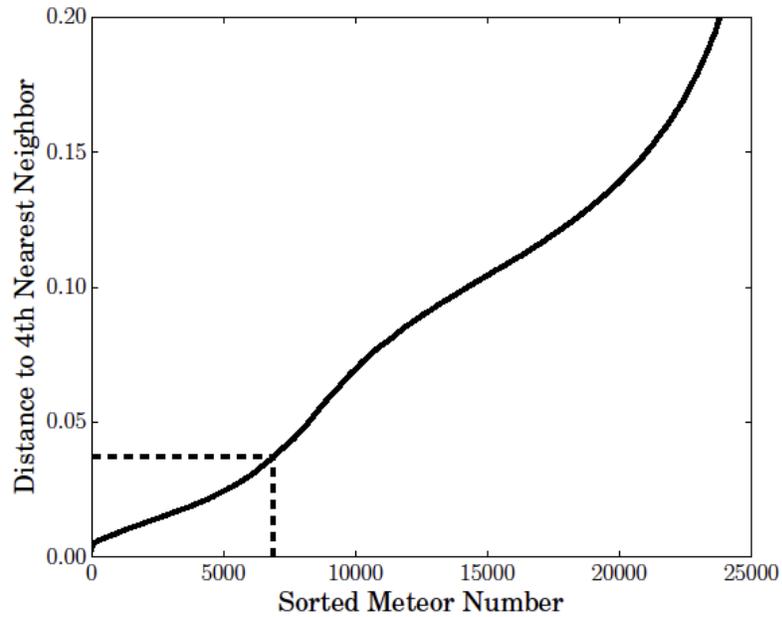

Fig. 1: A sorted (in ascending order) fourth nearest neighbor graph for the NASA and SOMN data set. The solid line shows the distance to the fourth nearest neighbor a given meteor. The dashed lines show the meteor with a fourth nearest neighbor distance greater than 23% of the data set. Setting ϵ to the distance value where the dashed lines and solid line (0.373) meet forces 23% of the meteors to be assigned to a cluster.

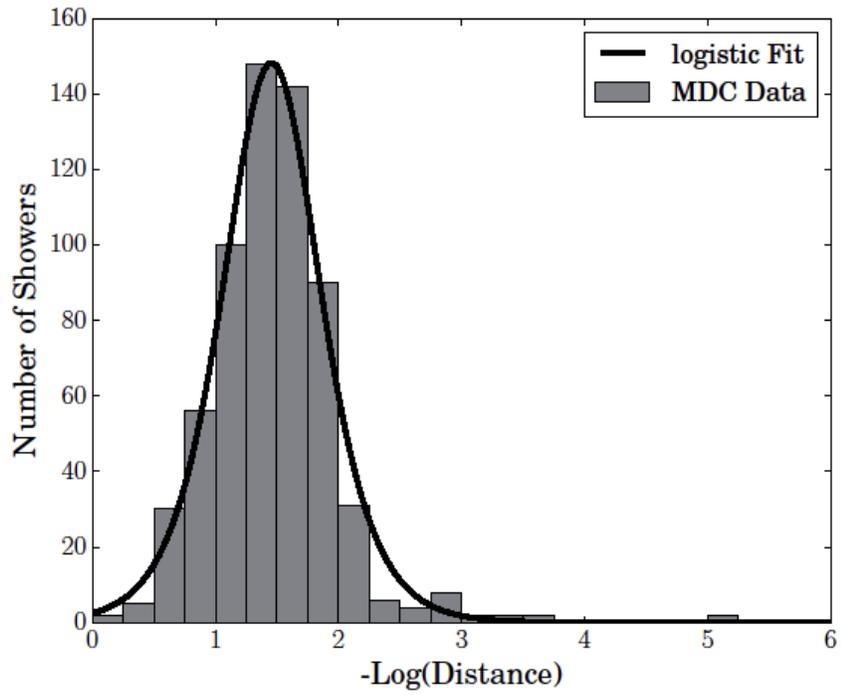

Fig. 2: A histogram of the negative log of the distance between each MDC shower and its nearest neighbor in our six-dimensional normalized space with the best-fit logistic distribution.

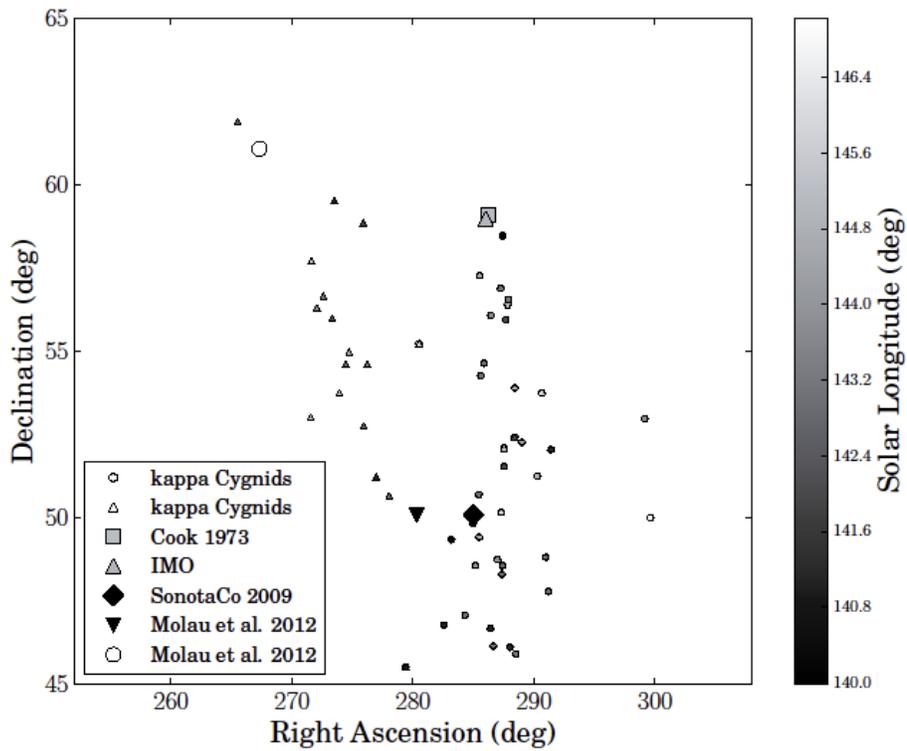

Fig. 3: The radiant and solar longitude of all meteors identified as members of either KCG cluster (small triangles and circles), as well as the mean values of 5 KCG entries reported in the MDC database.

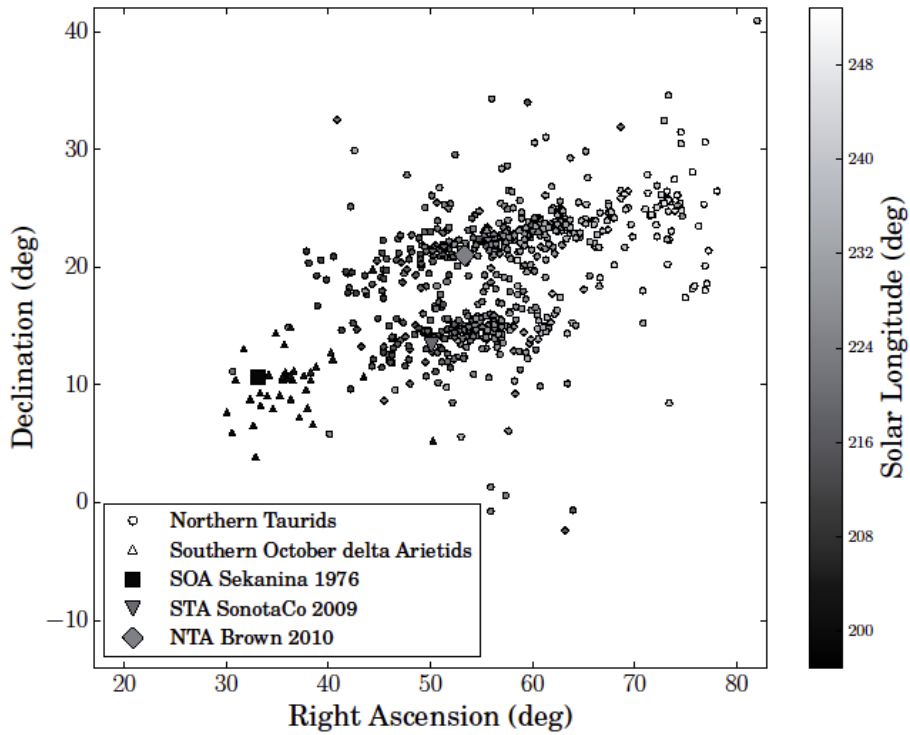

Fig. 4: The radiant and solar longitude of all meteors identified as either a NTA (circles) or SOA (triangles). MDC values of the SOA, NTA, and STA are shown for comparison. The Northern Taurid cluster is clearly the merged Southern and Northern Taurid showers.

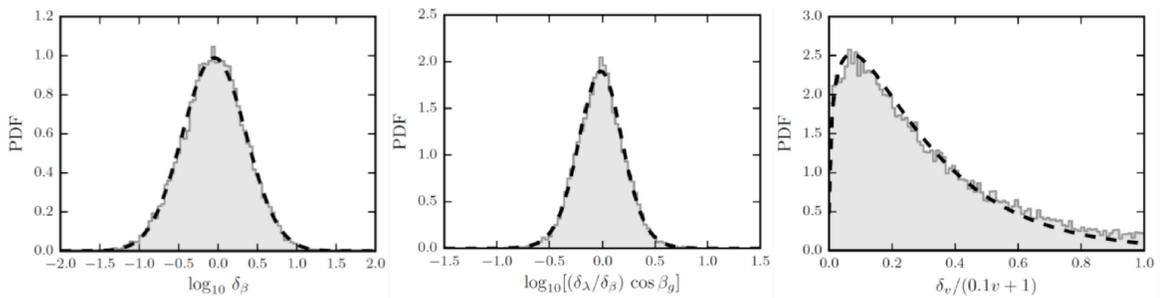

Fig. 5: Scaled histograms of errors for the NASA and SOMN data sets in gray, and their associated fitted probability density distributions used for generating the simulated sporadic meteor data. The left plot shows the logarithm of the error in geocentric latitude of the radiant. The middle plot shows the logarithm of the scaled error in geocentric longitude of the radiant. The right plot shows the scaled error in geocentric velocity magnitude.

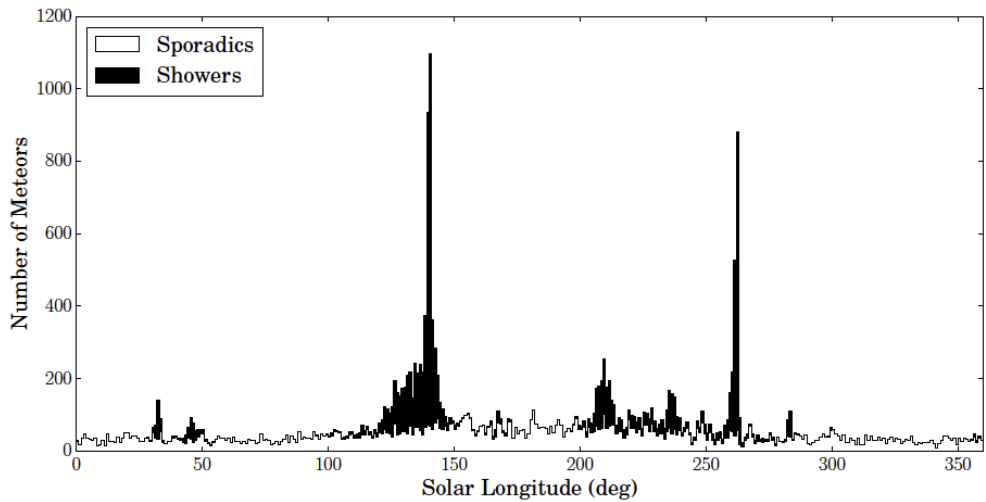

Fig. 6: A stacked histogram of the number of sporadic and shower meteors as a function of solar longitude. A meteor that was assigned to a cluster at least once out of the 1000 iterations is classified as a shower meteor. The bin size is one degree. The lack of peaks or valleys during periods of high shower activity is evidence for few false positives and negatives.

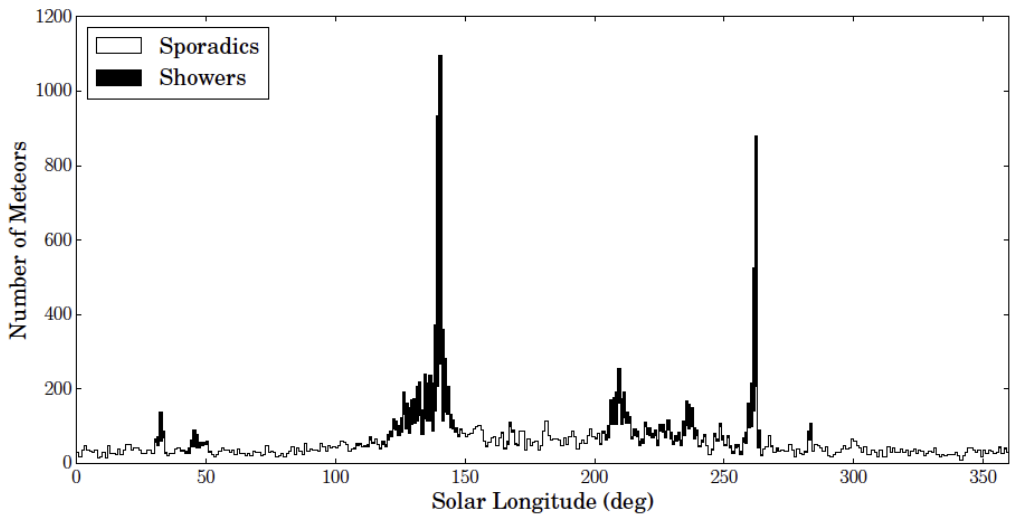

Fig. 7: A stacked histogram of the weighted number of sporadic and shower meteors as a function of solar longitude. A meteor will contribute to a shower bin the fraction of times it was classified as a shower member in the 1000 DBSCAN iterations, and to a sporadic bin the fraction of times it was not classified as a shower member. The bin size is one degree. The large peaks in the sporadic distributions during periods of high shower activity are evidence of false negatives.

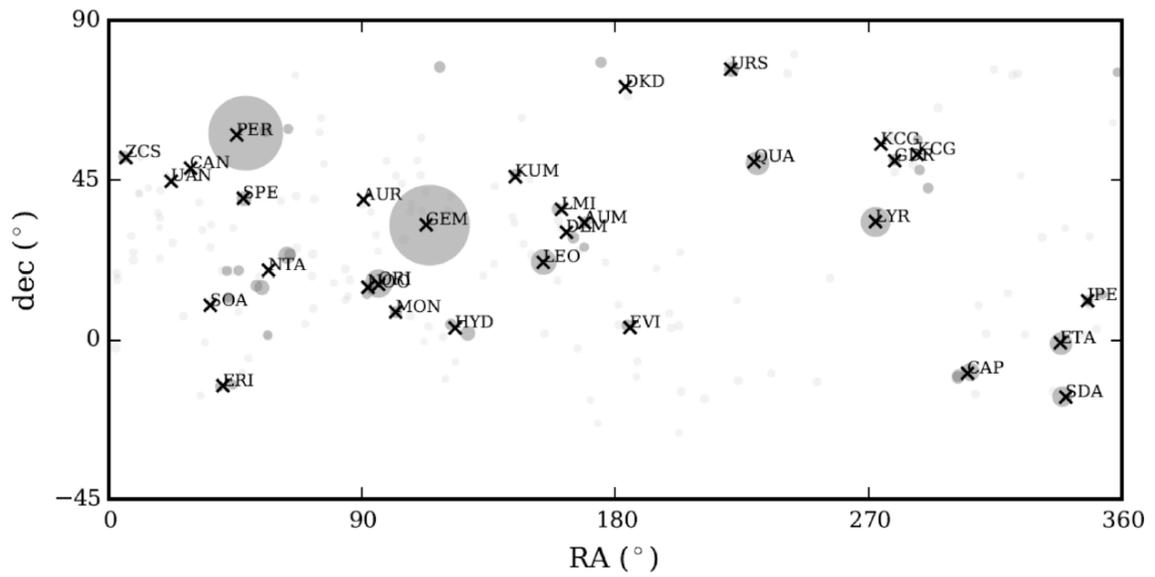

Fig. 8: The radiant location of showers found with DBSCAN (x symbols) and the wavelet method (circles). The size of the circle corresponds to the size of the shower's wavelet coefficient. Light gray circles are showers that are below the 20 wavelet coefficient threshold.

# Tables

| Shower | Confidence | Detected | N | $\lambda_\odot$ (deg) | RA (deg) | Dec (deg) | $v_g$ (km/s) |
|---|---|---|---|---|---|---|---|
| Perseids (PER) | 0.9999 | 1000 | 4346 | 138.5 (4.0) | 45.4 (6.8) | 57.7 (2.1) | 57.9 (2.6) |
| Geminids (GEM) | 1.0000 | 1000 | 1584 | 261.8 (1.1) | 112.8 (2.3) | 32.3 (1.7) | 33.3 (1.7) |
| Orionids (ORI) | 0.9995 | 1000 | 977 | 209.2 (2.7) | 95.9 (2.5) | 15.7 (1.6) | 64.1 (2.6) |
| Northern Taurids (NTA) | 0.9904 | 1000 | 615 | 227.7 (7.5) | 56.7 (6.0) | 19.5 (4.3) | 27.3 (1.7) |
| Leonids (LEO) | 0.9991 | 1000 | 463 | 236.4 (2.2) | 154.4 (2.1) | 21.7 (1.6) | 67.8 (2.1) |
| Southern delta Aquariids (SDA) | 0.9998 | 1000 | 345 | 128.1 (3.1) | 340.0 (2.8) | -16.0 (1.9) | 39.5 (1.5) |
| eta Aquariids (ETA) | 1.0000 | 1000 | 280 | 46.0 (2.2) | 338.0 (1.8) | -0.8 (1.3) | 64.6 (1.7) |
| alpha Capricornids (CAP) | 1.0000 | 1000 | 284 | 125.8 (4.1) | 305.0 (2.4) | -9.4 (1.7) | 22.3 (1.4) |
| April Lyrids (LYR) | 1.0000 | 1000 | 204 | 32.4 (0.8) | 272.3 (1.8) | 33.2 (1.3) | 46.2 (1.9) |
| sigma Hydrids (HYD) | 0.9997 | 1000 | 187 | 253.8 (4.9) | 123.1 (4.0) | 3.4 (1.3) | 58.4 (1.7) |
| Quadrantids (QUA) | 0.9999 | 1000 | 139 | 283.2 (0.5) | 229.4 (2.9) | 50.1 (1.6) | 39.5 (1.5) |
| September epsilon Perseids (SPE) | 1.0000 | 1000 | 127 | 167.6 (1.5) | 47.8 (2.3) | 39.8 (1.3) | 63.1 (2.3) |
| December Monocerotids (MON) | 0.9998 | 1000 | 100 | 260.6 (1.4) | 101.9 (1.4) | 7.8 (1.4) | 40.5 (1.1) |
| zeta Cassiopeiids (ZCS) | 0.9999 | 1000 | 68 | 113.1 (1.2) | 6.0 (2.7) | 51.3 (1.2) | 56.4 (1.7) |
| July Pegasids (JPE) | 0.9999 | 979 | 66 | 108.6 (1.6) | 347.7 (1.6) | 11.1 (0.9) | 63.7 (1.7) |
| Ursids (URS) | 0.9999 | 1000 | 20 | 270.5 (0.5) | 221.1 (3.6) | 76.3 (0.9) | 32.5 (0.8) |
| eta Virginids (EVI) | 0.9852 | 993 | 31 | 357.1 (0.9) | 185.3 (1.3) | 3.5 (1.2) | 26.4 (1.0) |
| eta Eridanids (ERI) | 0.9709 | 907 | 24 | 132.4 (0.7) | 39.2 (1.1) | -13.3 (0.9) | 63.5 (0.8) |
| December Leonis Minorids (DLM) | 0.9992 | 655 | 127 | 269.3 (3.3) | 162.6 (3.1) | 30.3 (2.0) | 62.2 (2.1) |
| Leonis Minorids (LMI) | 0.9997 | 761 | 45 | 210.2 (1.3) | 160.9 (1.9) | 36.7 (1.0) | 60.7 (2.3) |
| Southern October delta Arietids (SOA) | 0.9855 | 581 | 39 | 201.4 (1.3) | 36.1 (2.1) | 9.8 (1.3) | 29.1 (1.2) |
| kappa Ursae Majorids (KUM) | 0.9962 | 612 | 29 | 222.8 (0.6) | 144.6 (2.0) | 45.9 (0.8) | 63.1 (1.2) |
| July Gamma Draconids (GDR) | 1.0000 | 555 | 14 | 125.7 (0.4) | 279.1 (1.3) | 50.4 (1.0) | 26.8 (1.2) |
| November Orionids (NOO) | 0.9999 | 447 | 26 | 249.5 (1.1) | 92.2 (1.3) | 15.0 (1.2) | 41.5 (1.1) |
| Aurigids (AUR) | 0.9998 | 415 | 20 | 158.4 (1.2) | 90.5 (1.2) | 39.4 (0.9) | 64.6 (1.2) |
| kappa Cygnids (KCG1) | 0.9974 | 311 | 39 | 143.8 (1.1) | 287.1 (2.8) | 52.2 (3.5) | 23.2 (1.3) |
| January xi Ursae Majorids (XUM) | 0.9995 | 347 | 16 | 299.4 (0.6) | 169.1 (1.8) | 32.9 (1.1) | 41.2 (0.9) |
| December kappa Draconids (DKD) | 0.9995 | 220 | 24 | 250.6 (1.0) | 183.3 (3.7) | 71.2 (2.5) | 43.0 (1.9) |
| kappa Cygnids (KCG2) | 0.9181 | 125 | 15 | 143.4 (1.0) | 273.9 (1.6) | 55.3 (1.4) | 20.2 (0.6) |
| upsilon Andromedids (UAN) | 0.9894 | 129 | 31 | 101.5 (1.1) | 22.0 (2.4) | 44.6 (1.3) | 57.1 (1.7) |
| c Andromedids (CAN) | 0.9978 | 109 | 11 | 108.1 (1.0) | 29.1 (1.7) | 48.2 (1.0) | 57.0 (1.1) |

Table 1: Detected showers in the NASA and SOMN data.

| Shower | Confidence | Detected | N | $\lambda_\odot$ (deg) | RA (deg) | Dec (deg) | Vel (km/s) |
|---|---|---|---|---|---|---|---|
| sigma Capricornids | 0.7623 | 330 | 32 | 119.6 (1.1) | 321.8 (1.7) | -11.5 (1.0) | 29.1 (1.6) |
| nu Cygnids | 0.5909 | 222 | 37 | 22.7 (0.7) | 290.5 (1.4) | 33.1 (1.5) | 39.0 (1.7) |
| April chi Librids | 0.9639 | 107 | 54 | 43.4 (1.1) | 242.3 (3.0) | -19.5 (2.0) | 30.3 (2.2) |

Table 2: Detected showers in the simulated sporadic data.